\documentclass[conference,final]{IEEEtran}
\IEEEoverridecommandlockouts

\usepackage[utf8]{inputenc}
\usepackage[T1]{fontenc}
\usepackage{microtype,newtxtext,newtxmath} % use newpx[text|math] for palatino instead of times
\usepackage{mathtools,gensymb,eurosym,url,soul} % provides amsmath, generic symbols, URL management and strikeout
\usepackage[font=small]{caption}
\usepackage{longtable,booktabs,array}

\usepackage[switch]{lineno}
\usepackage{acronym}
\usepackage{graphicx}
\usepackage{xcolor}%[usenames,dvipsnames,svgnames,table]{xcolor}
\usepackage{xr} % for referring to the paper document
\usepackage{textcomp}
\usepackage{hyperref}
\hypersetup{colorlinks=true, breaklinks=true, %pagebackref=true,
  urlcolor=blue, linkcolor=blue,anchorcolor=blue,citecolor=blue,
  hypertexnames=true, final=true,
  pdfpagemode = UseNone, %FullScreen, %UseThumbs, %UseOutlines,
  pdfauthor = {},
  pdftitle = {},
  pdfsubject = {},
  pdfkeywords = {}
}

\usepackage{amsmath}
\usepackage{array}
\usepackage{footnote}
\usepackage{amsmath}

\graphicspath{{figures/} {img/} {./}}
\DeclareGraphicsExtensions{.pdf,.png,.jpg,.mps}

\IEEEdisplaynontitleabstractindextext

\begin{document}
\title{ElfCore: A 28nm Neural Processor Enabling Dynamic Structured Sparse Training and Online Self-Supervised Learning with Activity-Dependent Weight Update\\
}
\acrodef{ADC}[ADC]{Analog-to-Digital Converter}
\acrodef{ADEXP}[AdExp-IF]{Adaptive Exponential Integrate-and-Fire}
\acrodef{ADM}[ADM]{Asynchronous Delta Modulator}
\acrodef{AE}[AE]{Address-Event}
\acrodef{AER}[AER]{Address-Event Representation}
\acrodef{AEX}[AEX]{AER EXtension board}
\acrodef{AFE}[AFE]{Analog Front-End}
\acrodef{AFM}[AFM]{Atomic Force Microscope}
\acrodef{AGC}[AGC]{Automatic Gain Control}
\acrodef{AI}[AI]{Artificial Intelligence}
\acrodef{AMDA}[AMDA]{AER Motherboard with D/A converters}
\acrodef{AMPA}[AMPA]{$\alpha$-Amino-3-hydroxy-5-methyl-4-isoxazolepropionic Acid}
\acrodef{ANN}[ANN]{Artificial Neural Network}
\acrodef{API}[API]{Application Programming Interface}
\acrodef{APMOM}[APMOM]{Alternate Polarity Metal On Metal}
\acrodef{ARM}[ARM]{Advanced RISC Machine}
\acrodef{ASIC}[ASIC]{Application Specific Integrated Circuit}
\acrodef{BCM}[BMC]{Bienenstock-Cooper-Munro}
\acrodef{BD}[BD]{Bundled Data}
\acrodef{BEOL}[BEOL]{Back-end of Line}
\acrodef{BG}[BG]{Bias Generator}
\acrodef{BMI}[BMI]{Brain-Machince Interface}
\acrodef{BTB}[BTB]{Band-to-Band tunnelling}
\acrodef{CA}[CA]{Cortical Automaton}
\acrodef{CAD}[CAD]{Computer Aided Design}
\acrodef{CAM}[CAM]{Content Addressable Memory}
\acrodef{CAVIAR}[CAVIAR]{Convolution AER Vision Architecture for Real-Time}
\acrodef{CCN}[CCN]{Cooperative and Competitive Network}
\acrodef{CDR}[CDR]{Clock-Data Recovery}
\acrodef{CFC}[CFC]{Current to Frequency Converter}
\acrodef{CHP}[CHP]{Communicating Hardware Processes}
\acrodef{CMIM}[CMIM]{Metal-Insulator-Metal Capacitor}
\acrodef{CML}[CML]{Current Mode Logic}
\acrodef{CMOL}[CMOL]{Hybrid CMOS nanoelectronic circuits}
\acrodef{CMOS}[CMOS]{Complementary Metal-Oxide-Semiconductor}
\acrodef{CNN}[CNN]{Convolutional Neural Network}
\acrodef{CNS}[CNS]{central Nervous System}
\acrodef{COTS}[COTS]{Commercial Off-The-Shelf}
\acrodef{CPG}[CPG]{Central Pattern Generator}
\acrodef{CPLD}[CPLD]{Complex Programmable Logic Device}
\acrodef{CPU}[CPU]{Central Processing Unit}
\acrodef{CSM}[CSM]{Cortical State Machine}
\acrodef{CSP}[CSP]{Constraint Satisfaction Problem}
\acrodef{CTXCTL}[CTXCTL]{CortexControl}
\acrodef{CV}[CV]{Coefficient of Variation}
\acrodef{DAC}[DAC]{Digital to Analog Converter}
\acrodef{DAS}[DAS]{Dynamic Auditory Sensor}
\acrodef{DAVIS}[DAVIS]{Dynamic and Active Pixel Vision Sensor}
\acrodef{DBN}[DBN]{Deep Belief Network}
\acrodef{DBS}[DBS]{Deep Brain Stimulation}
\acrodef{DFA}[DFA]{Deterministic Finite Automaton}
\acrodef{DIBL}[DIBL]{Drain-Induced Barrier-Lowering}
\acrodef{DI}[DI]{Delay Insensitive}
\acrodef{divmod3}[DIVMOD3]{Divisibility of a number by three}
\acrodef{DMA}[DMA]{Direct Memory Access}
\acrodef{DNF}[DNF]{Dynamic Neural Field}
\acrodef{DNN}[DNN]{Deep Neural Network}
\acrodef{DOF}[DOF]{Degrees of Freedom}
\acrodef{DPE}[DPE]{Dynamic Parameter Estimation}
\acrodef{DPI}[DPI]{Differential Pair Integrator}
\acrodef{DRAM}[DRAM]{Dynamic Random Access Memory}
\acrodef{DR}[DR]{Dual Rail}
\acrodef{DRRZ}[DR-RZ]{Dual-Rail Return-to-Zero}
\acrodef{DSP}[DSP]{Digital Signal Processor}
\acrodef{DVS}[DVS]{Dynamic Vision Sensor}
\acrodef{DYNAP}[DYNAP]{Dynamic Neuromorphic Asynchronous Processor}
\acrodef{EBL}[EBL]{Electron Beam Lithography}
\acrodef{ECG}[ECG]{Electrocardiography}
\acrodef{ECoG}[ECoG]{Electrocorticography}
\acrodef{EDVAC}[EDVAC]{Electronic Discrete Variable Automatic Computer}
\acrodef{EEG}[EEG]{Electroencephalography}
\acrodef{EIN}[EIN]{Excitatory-Inhibitory Network}
\acrodef{EM}[EM]{Expectation Maximization}
\acrodef{EMG}[EMG]{Electromyography}
\acrodef{EOG}[EOG]{Electrooculogram}
\acrodef{EPSC}[EPSC]{Excitatory Post-Synaptic Current}
\acrodef{EPSP}[EPSP]{Excitatory Post-Synaptic Potential}
\acrodef{EZ}[EZ]{Epileptogenic Zone}
\acrodef{FDSOI}[FDSOI]{Fully-Depleted Silicon on Insulator}
\acrodef{FET}[FET]{Field-Effect Transistor}
\acrodef{FFT}[FFT]{Fast Fourier Transform}
\acrodef{FI}[F-I]{Frequency--Current}
\acrodef{FMA}[FMA]{Floating Microelectrode Array} 
\acrodef{FNN}[FNN]{Feed-forward Neural Network}
\acrodef{FPGA}[FPGA]{Field Programmable Gate Array}
\acrodef{FR}[FR]{Fast Ripple}
\acrodef{FSA}[FSA]{Finite State Automaton}
\acrodef{FSM}[FSM]{Finite State Machine}
\acrodef{GABA}[GABA]{$\gamma$-Aminobutanoic Acid}
\acrodef{GIDL}[GIDL]{Gate-Induced Drain Leakage}
\acrodef{GOPS}[GOPS]{Giga-Operations per Second}
\acrodef{GPIO}[GPIO]{General Purpose I/O}
\acrodef{GPU}[GPU]{Graphical Processing Unit}
\acrodef{GT}[GT]{Ground Truth}
\acrodef{GUI}[GUI]{Graphical User Interface}
\acrodef{HAL}[HAL]{Hardware Abstraction Layer}
\acrodef{HFO}[HFO]{High Frequency Oscillation}
\acrodef{HH}[H\&H]{Hodgkin \& Huxley}
\acrodef{HMM}[HMM]{Hidden Markov Model}
\acrodef{HR}[HR]{Human Readable}
\acrodef{HRS}[HRS]{High-Resistive State}
\acrodef{HSE}[HSE]{Handshaking Expansion}
\acrodef{HW}[HW]{Hardware}
\acrodef{hWTA}[hWTA]{Hard Winner-Take-All}
\acrodef{IC}[IC]{Integrated Circuit}
\acrodef{ICT}[ICT]{Information and Communication Technology}
\acrodef{iEEG}[iEEG]{Intracranial Electroencephalography}
\acrodef{IF2DWTA}[IF2DWTA]{Integrate \& Fire 2-Dimensional WTA}
\acrodef{IF}[I\&F]{Integrate-and-Fire}
\acrodef{IFSLWTA}[IFSLWTA]{Integrate \& Fire Stop Learning WTA}
\acrodef{IMU}[IMU]{Inertial Measurement Unit}
\acrodef{INCF}[INCF]{International Neuroinformatics Coordinating Facility}
\acrodef{INI}[INI]{Institute of Neuroinformatics}
\acrodef{IO}[I/O]{Input/Output}
\acrodef{IoT}[IoT]{Internet of Things}
\acrodef{IP}[IP]{Intellectual Property}
\acrodef{IPSC}[IPSC]{Inhibitory Post-Synaptic Current}
\acrodef{IPSP}[IPSP]{Inhibitory Post-Synaptic Potential}
\acrodef{ISI}[ISI]{Inter-Spike Interval}
\acrodef{JFLAP}[JFLAP]{Java - Formal Languages and Automata Package}
\acrodef{LEDR}[LEDR]{Level-Encoded Dual-Rail}
\acrodef{LFP}[LFP]{Local Field Potential}
\acrodef{LIFE}[LIFE]{Longitudinal Intrafascicular Electrodes}
\acrodef{LIF}[LI\&F]{Leaky Integrate-and-Fire}
\acrodef{LLC}[LLC]{Low Leakage Cell}
\acrodef{LNA}[LNA]{Low-Noise Amplifier}
\acrodef{LPF}[LPF]{Low Pass Filter}
\acrodef{LRS}[LRS]{Low-Resistive State}
\acrodef{LSM}[LSM]{Liquid State Machine}
\acrodef{LTD}[LTD]{Long Term Depression}
\acrodef{LTI}[LTI]{Linear Time-Invariant}
\acrodef{LTP}[LTP]{Long Term Potentiation}
\acrodef{LTU}[LTU]{Linear Threshold Unit}
\acrodef{LUT}[LUT]{Look-Up Table}
\acrodef{LVDS}[LVDS]{Low Voltage Differential Signaling}
\acrodef{MCMC}[MCMC]{Markov-Chain Monte Carlo}
\acrodef{MEA}[MEA]{Multielectrode Arrays}
\acrodef{MEMS}[MEMS]{Micro Electro Mechanical System}
\acrodef{MFR}[MFR]{Mean Firing Rate}
\acrodef{MIM}[MIM]{Metal Insulator Metal}
\acrodef{ML}[ML]{Machine Learning}
\acrodef{MLP}[MLP]{Multilayer Perceptron}
\acrodef{MOSCAP}[MOSCAP]{Metal Oxide Semiconductor Capacitor}
\acrodef{MOSFET}[MOSFET]{Metal Oxide Semiconductor Field-Effect Transistor}
\acrodef{MOS}[MOS]{Metal Oxide Semiconductor}
\acrodef{MRI}[MRI]{Magnetic Resonance Imaging}
\acrodef{NCS}[NCS]{Neuromorphic Cognitive Systems}
\acrodef{NDFSM}[NDFSM]{Non-deterministic Finite State Machine} 
\acrodef{ND}[ND]{Noise-Driven}
\acrodef{NEF}[NEF]{Neural Engineering Framework}
\acrodef{NHML}[NHML]{Neuromorphic Hardware Mark-up Language}
\acrodef{NIL}[NIL]{Nano-Imprint Lithography}
\acrodef{NI}[NI]{Neural Interface}
\acrodef{NMDA}[NMDA]{\textit{N}-Methyl-\textsc{d}-aspartate}
\acrodef{NME}[NE]{Neuromorphic Engineering}
\acrodef{NN}[NN]{Neural Network}
\acrodef{NOC}[NoC]{Network-on-Chip}
\acrodef{NRZ}[NRZ]{Non-Return-to-Zero}
\acrodef{NSM}[NSM]{Neural State Machine}
\acrodef{OR}[OR]{Operating Room}
\acrodef{OTA}[OTA]{Operational Transconductance Amplifier}
\acrodef{PCB}[PCB]{Printed Circuit Board}
\acrodef{PCHB}[PCHB]{Pre-Charge Half-Buffer}
\acrodef{PCM}[PCM]{Phase Change Memory}
\acrodef{PC}[PC]{Personal Computer}
\acrodef{PDK}[PDK]{Process Design Kit}
\acrodef{PE}[PE]{Phase Encoding}
\acrodef{PFA}[PFA]{Probabilistic Finite Automaton}
\acrodef{PFC}[PFC]{Prefrontal Cortex}
\acrodef{PFM}[PFM]{Pulse Frequency Modulation}
\acrodef{PNI}[PNI]{Peripheral Nerve Interface}
\acrodef{PNS}[PNS]{Peripheral Nervous System}
\acrodef{PPG}[PPG]{Photoplethysmography}
\acrodef{PR}[PR]{Production Rule}
\acrodef{PSC}[PSC]{Post-Synaptic Current}
\acrodef{PSP}[PSP]{Post-Synaptic Potential}
\acrodef{PSTH}[PSTH]{Peri-Stimulus Time Histogram}
\acrodef{PV}[PV]{Parvalbumin}
\acrodef{QDI}[QDI]{Quasi Delay Insensitive}
\acrodef{RAM}[RAM]{Random Access Memory}
\acrodef{RA}[RA]{Resected Area}
\acrodef{RDF}[RDF]{Random Dopant Fluctuation}
\acrodef{RELU}[ReLu]{Rectified Linear Unit}
\acrodef{RLS}[RLS]{Recursive Least-Squares}
\acrodef{RMSE}[RMSE]{Root Mean Square-Error}
\acrodef{RMS}[RMS]{Root Mean Square}
\acrodef{RNN}[RNN]{Recurrent Neural Network}
\acrodef{ROLLS}[ROLLS]{Reconfigurable On-Line Learning Spiking}
\acrodef{RRAM}[R-RAM]{Resistive Random Access Memory}
\acrodef{R}[R]{Ripple}
\acrodef{RISC}[RISC]{Reduced Instruction Set Computer}
\acrodef{RSA}[RSA]{Respiratory Sinus Arrhythmia}
\acrodef{SAC}[SAC]{Selective Attention Chip}
\acrodef{SAT}[SAT]{Boolean Satisfiability Problem}
\acrodef{SCI}[SCI]{Spinal Cord Injury}
\acrodef{SCX}[SCX]{Silicon CorteX}
\acrodef{SD}[SD]{Signal-Driven}
\acrodef{SEM}[SEM]{Spike-based Expectation Maximization}
\acrodef{SLAM}[SLAM]{Simultaneous Localization and Mapping}
\acrodef{SNN}[SNN]{Spiking Neural Network}
\acrodef{SNR}[SNR]{Signal to Noise Ratio}
\acrodef{SOC}[SoC]{System-On-Chip}
\acrodef{SOI}[SOI]{Silicon on Insulator}
\acrodef{SOZ}[SOZ]{Seizure Onset Zone}
\acrodef{SP}[SP]{Separation Property}
\acrodef{SPI}[SPI]{Serial Peripheral Interface}
\acrodef{SRAM}[SRAM]{Static Random Access Memory}
\acrodef{SST}[SST]{Somatostatin}
\acrodef{STDP}[STDP]{Spike-Timing Dependent Plasticity}
\acrodef{STD}[STD]{Short-Term Depression}
\acrodef{STP}[STP]{Short-Term Plasticity}
\acrodef{STT-MRAM}[STT-MRAM]{Spin-Transfer Torque Magnetic Random Access Memory}
\acrodef{STT}[STT]{Spin-Transfer Torque}
\acrodef{SVM}[SVM]{Support Vector Machine}
\acrodef{SW}[SW]{Software}
\acrodef{sWTA}[sWTA]{soft Winner-Take-All}
\acrodef{TCAM}[TCAM]{Ternary Content-Addressable Memory}
\acrodef{TFT}[TFT]{Thin Film Transistor}
\acrodef{TIME}[TIME]{Transverse Intrafascicular Multichannel Electrode}
\acrodef{TLE}[TLE]{Temporal Lobe Epilepsy}
\acrodef{UEA}[UEA]{Utah Electrode Array}
\acrodef{USB}[USB]{Universal Serial Bus}
\acrodef{USEA}[USEA]{Utah Slanted Electrode Array}
\acrodef{VHDL}[VHDL]{VHSIC Hardware Description Language}
\acrodef{VHSIC}[VHSIC]{Very High Speed Integrated Circuits}
\acrodef{VIP}[VIP]{Vasoactive Intestinal Peptide}
\acrodef{VLSI}[VLSI]{Very Large Scale Integration}
\acrodef{VNS}[VNS]{Vagal Nerve Stimulation}
\acrodef{VOR}[VOR]{Vestibulo-Ocular Reflex}
\acrodef{VSA}[VSA]{Vector Symbolic Architecture}
\acrodef{WCST}[WCST]{Wisconsin Card Sorting Test}
\acrodef{WTA}[WTA]{Winner-Take-All}
\acrodef{XML}[XML]{eXtensible Mark-up Language}

\author{
Zhe Su and Giacomo Indiveri \\
\IEEEauthorblockA{\textit{Institute of Neuroinformatics} \textit{University of Zurich and ETH Zurich}} zhesu@ini.ethz.ch
}

\maketitle

\begin{abstract}
In this paper, we present ElfCore, a 28nm digital spiking neural network processor tailored for event-driven sensory signal processing. ElfCore is the \textit{first} to efficiently integrate: (1) a local online self-supervised learning engine that enables multi-layer temporal learning without labeled inputs; (2) a dynamic structured sparse training engine that supports high-accuracy sparse-to-sparse learning; and (3) an activity-dependent sparse weight update mechanism that selectively updates weights based solely on input activity and network dynamics. Demonstrated on tasks including gesture recognition, speech, and biomedical signal processing, ElfCore outperforms state-of-the-art solutions with up to 16$\times$ lower power consumption, 3.8$\times$ reduced on-chip memory requirements, and 5.9$\times$ greater network capacity efficiency.
\end{abstract}

\begin{IEEEkeywords}
self-supervised learning; dynamic structured sparse training; sparse weight update
\end{IEEEkeywords}

\section{Introduction}
\label{sec:intro}

\begin{figure*}
\centering
 \includegraphics[width=0.99\linewidth]{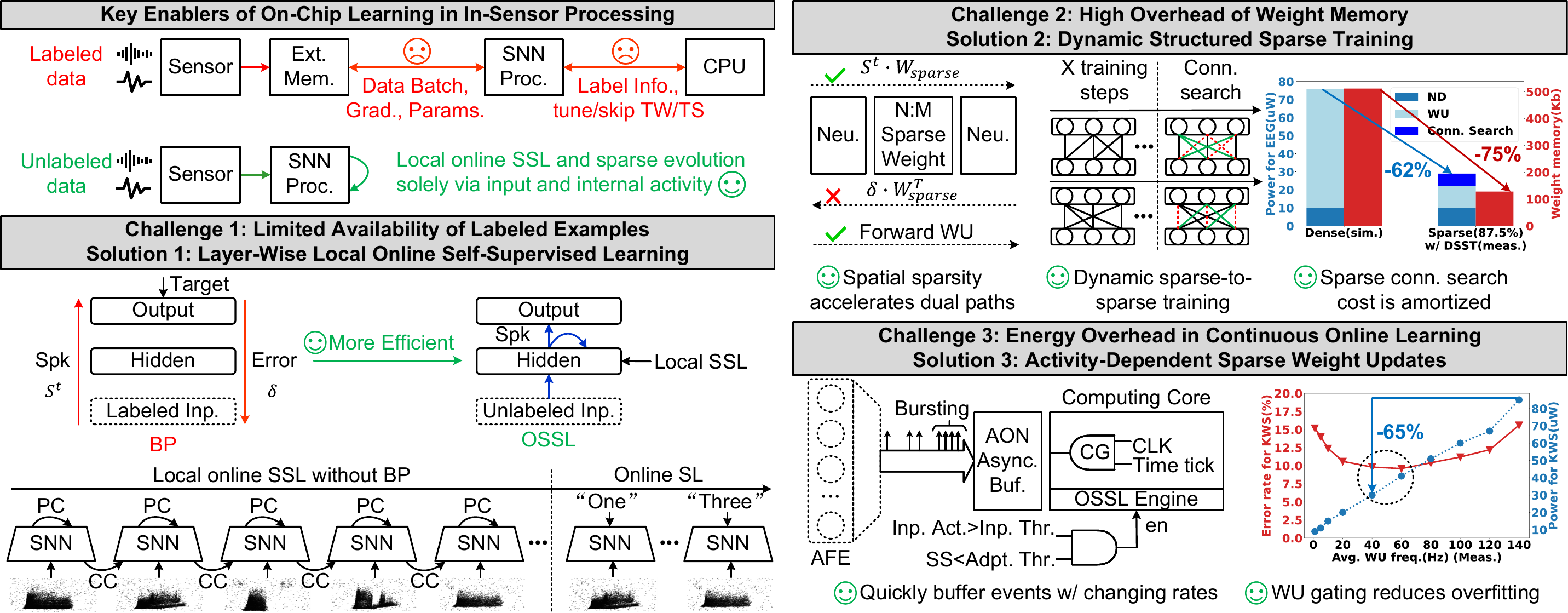}
\caption{Requirements for on-chip learning in processing streaming event data, highlighting three key challenges and their corresponding solutions.}
\label{fig:intro}
\vskip -0.2cm
\end{figure*}

Spiking neural network (SNN) processors offer reduced bandwidth requirement and power consumption while enabling event-driven processing on demand. Their on-chip multi-layer learning lets edge devices adapt to the shifting input distribution, overcoming output-layer few-shot learning’s limitations in adjusting underlying features. Prior work has focused on either unsupervised techniques for static data \cite{vlsi_4096} or supervised (SL) methods \cite{vlsi_FSNAP, vlsi_anp, vlsi_reckon}. To address real-world applications where labeled data is scarce but unlabeled streaming data is abundant, a layer-wise local online self-supervised learning (OSSL) method was introduced. This approach integrates predictive coding (PC) within individual samples and contrastive coding (CC) across samples, thereby eliminating the need for labeled inputs (Fig. \ref{fig:intro}).

At the same time, a dynamic structured sparse training (DSST) process tackles limited on-chip weight memory resource challenges by periodically pruning and regrowing connections for efficient sparse-to-sparse training. This reduces memory requirements by up to 75\% with minimal energy overhead, as connection updates are much less frequent than OSSL. By integrating OSSL, which removes error backpropagation, with structured weight sparsity, the processor accelerates dual forward data paths, reducing power consumption by 62\%.

Finally, an activity-dependent sparse weight update (WU) mechanism uses input activity (IA) and a similarity score (SS) from neural dynamics (ND) to gate WU layer-wise, overcoming the limitations of traditional accuracy-driven methods, like time window (TW) tuning \cite{vlsi_FSNAP} and time step (TS) skipping \cite{vlsi_anp}, which rely on external schedulers and are unsuitable for streaming data. This reduces power by up to 65\%, lowers noise, and improves robustness by mitigating overfitting. Temporal sparsity is further enhanced by an always-on (AON) asynchronous SerDes, which adapts to sensory input rates and gates the core until the next TS arrives.

\section{Proposed Design}
\label{sec:proposed}

\subsection{Architecture Overview}

\begin{figure}
\centering
\includegraphics[width=0.99\linewidth]{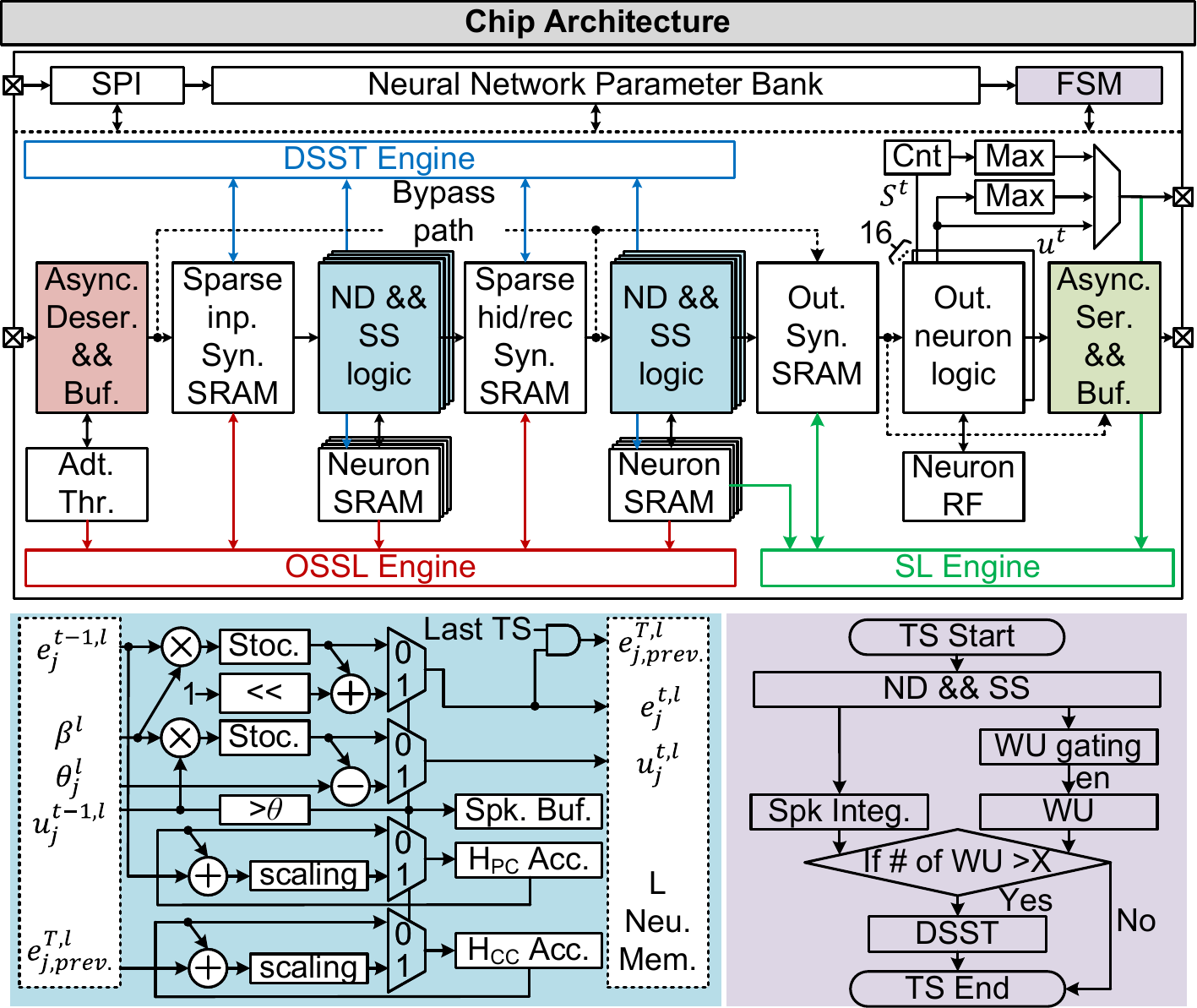}
\caption{Chip architecture (top), neuron dynamics and similarity score logic (bottom left), and FSM (bottom right).}
\label{fig:architecture}
\vskip -0.2cm
\end{figure}

The architecture features a two-hidden-layer network with bypass connections to the output (Fig. \ref{fig:architecture}). Each hidden layer contains four PEs that operate in parallel. Neuron SRAM stores spike traces across three TSs per neuron, supporting multi-timescale local learning: the current TS’s trace for WU, an earlier TS’s trace for PC, and the trace from the final TS of the previous sample for CC. The OSSL engine updates sparse weights, the DSST engine learns sparse connectivity, and the SL manages the output layer learning. Forward spike integration (SI) and WU run concurrently, and DSST is activated once enough WU cycles have completed. 

\subsection{Asynchronous SerDes Interface}

\begin{figure}
\centering
\includegraphics[width=0.99\linewidth]{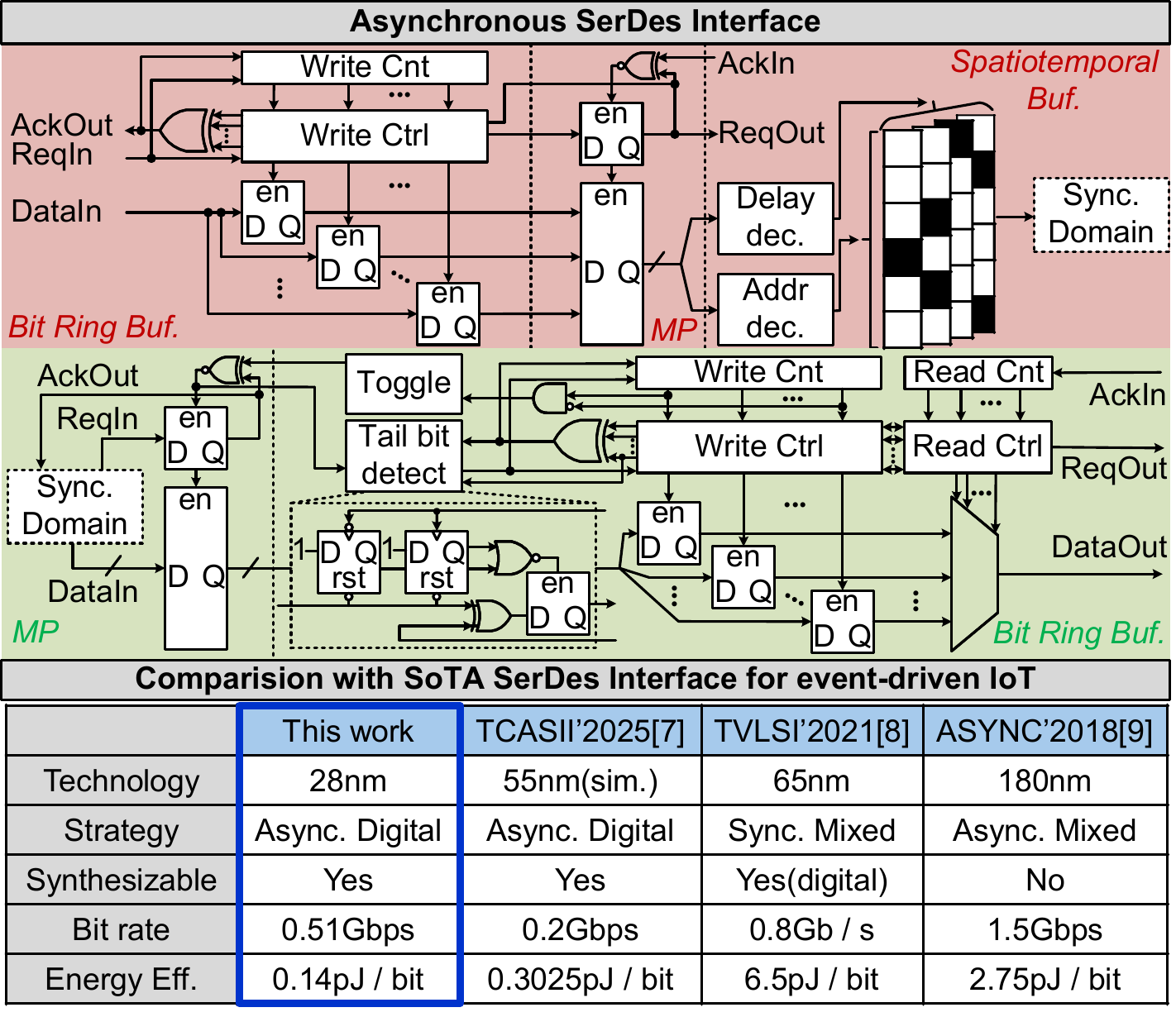}
\caption{Asynchronous de-serializer (top), asynchronous serializer (middle), and comparison with SoTA designs (bottom).}
\label{fig:serdes}
\vskip -0.2cm
\end{figure}

An asynchronous deserializer, which converts serial spike packets into 30-bit parallel packets, enabling the flexible input dimension (Fig. \ref{fig:serdes}). A spatiotemporal buffer stores 512-bit spike vectors with a 4-slot depth to emulate axonal delays, thereby enhancing temporal dynamics. The serializer supports inter-chip communication for deeper networks. Leveraging the Mousetrap (MP) pipeline \cite{vlsi_tabula, vlsi_jetcas} and ring structure, the asynchronous SerDes achieves 54\% better energy efficiency than SoTA solutions for event-driven IoT.
\nocite{vlsi_ning, vlsi_luca, vlsi_srrt}

\subsection{On-chip learning of both sparse weights and connectivity}

\begin{figure}
\centering
\includegraphics[width=0.99\linewidth]{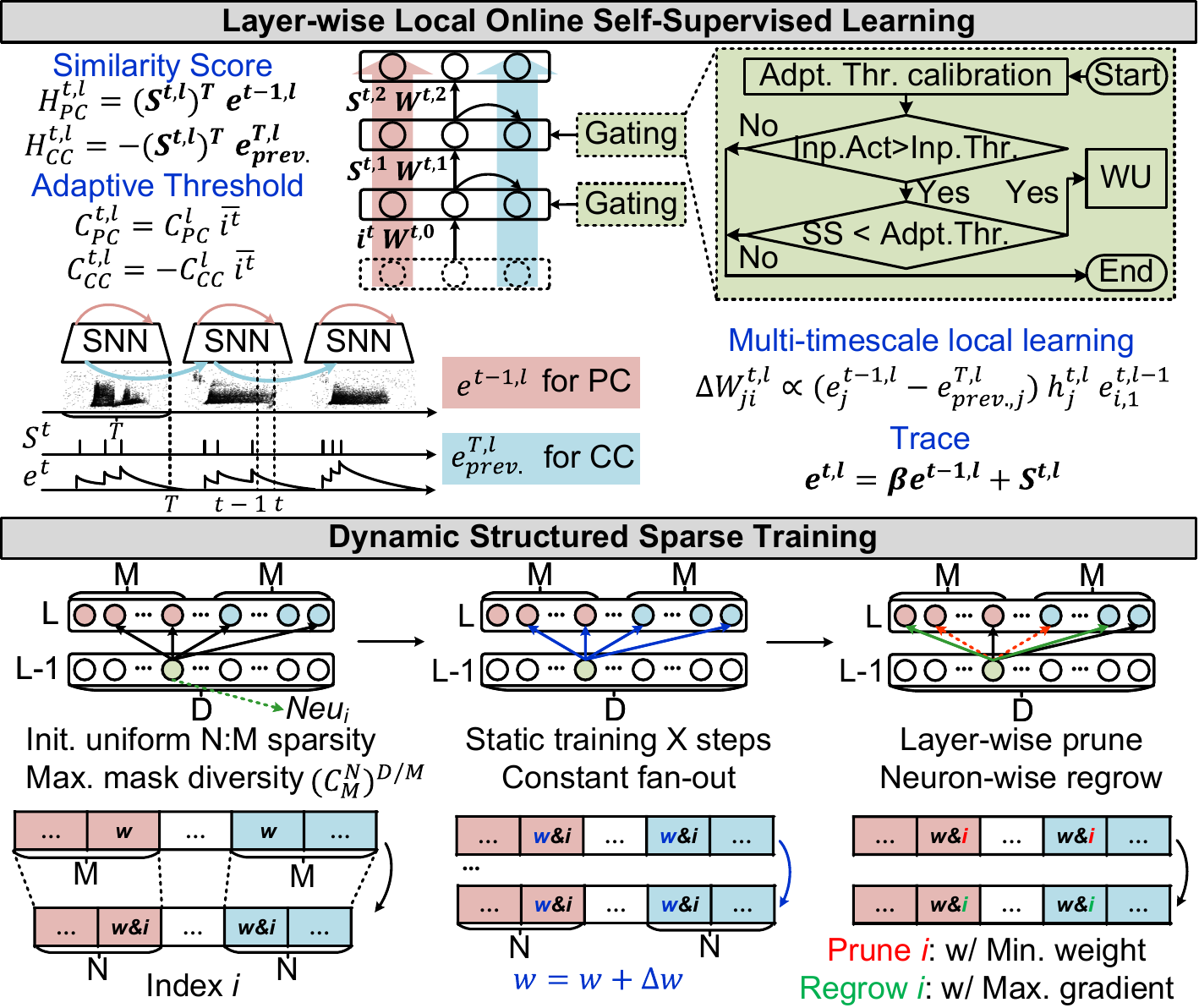}
\caption{
On-chip learning: Enhanced OSSL with simultaneous PC and CC (top), and DSST employing sparse-to-sparse training (bottom).
}
\label{fig:algorithm}
\end{figure}

ElfCore’s OSSL surpasses \cite{vlsi_espp, vlsi_clapp} by running PC and CC concurrently in every layer, removing the global class-transition flag. The only condition left—that consecutive samples are very likely drawn from different classes—is typically met in multi-class tasks (Fig. \ref{fig:algorithm}). Intrinsic layer-wise WU gating within OSSL is achieved by comparing IA with a global threshold, and SS with an adaptive layer-specific threshold. 

Unlike progressive pruning (dense-to-sparse), DSST starts directly with uniform N:M sparsity to maximize mask diversity. After synaptic weights are learned over multiple iterations, DSST prunes the k smallest weights between neuron layers and regrows an equal number of connections with the largest gradients, executed on an N:M group basis.

\subsection{Efficient dynamic structured sparse training}

\begin{figure}
\centering
\includegraphics[width=0.99\linewidth]{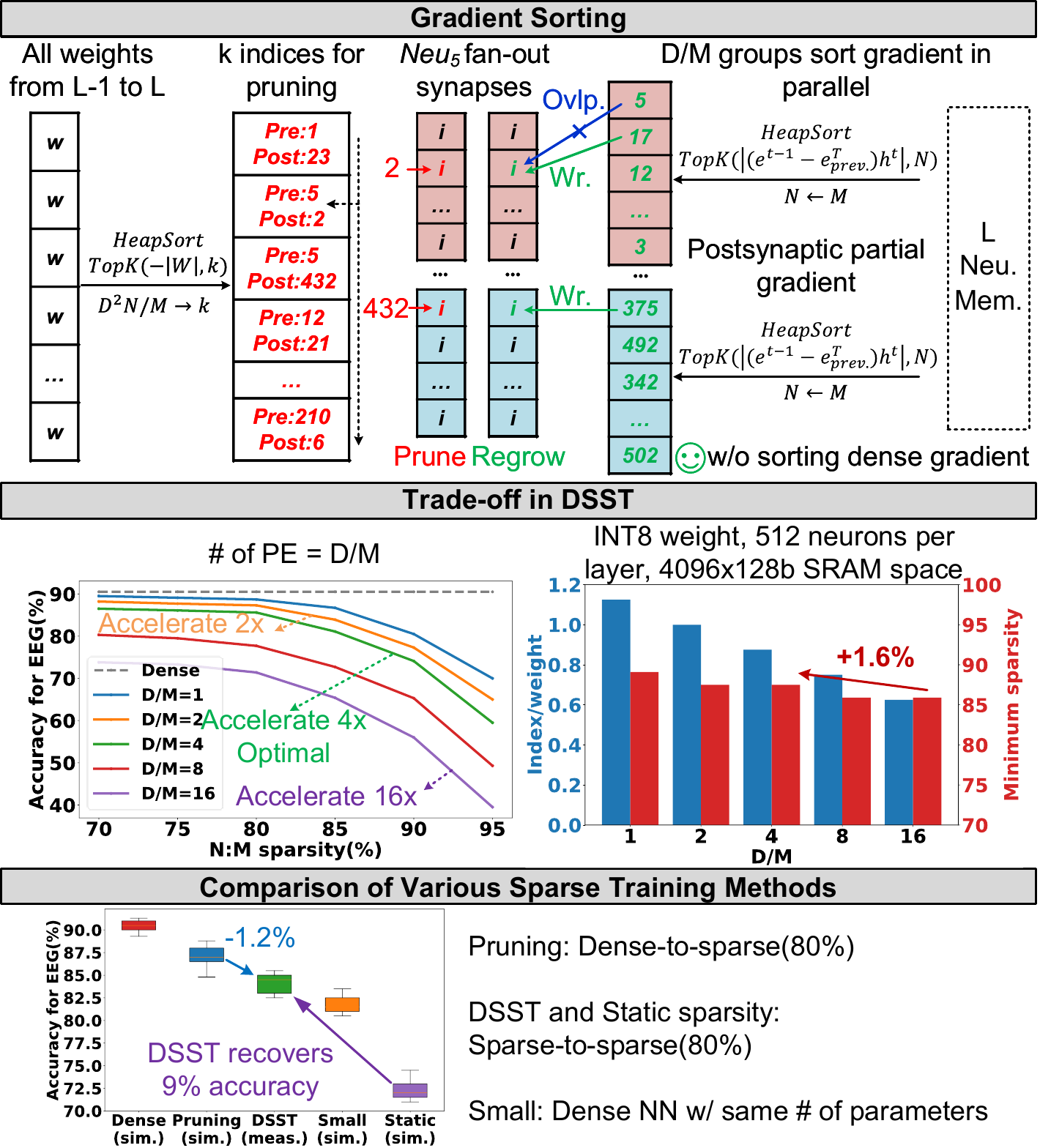}
\caption{
Enhanced efficiency in DSST: Avoids dense gradient sorting (top). The trade-off between acceleration and accuracy (middle). DSST restores accuracy (bottom).
}
\label{fig:dsst}
\vskip -0.2cm
\end{figure}

DSST improves on \cite{vlsi_rigl} by replacing dense gradient sorting with a novel method that separates pre- and post-gradient components (Fig. \ref{fig:dsst}). This enables efficient reuse of post-gradient sorting across presynaptic neurons, as pre-gradients are shared across all fan-out connections of a neuron, reducing sorting complexity from the synapse to the neuron level. To strike a balance between mask diversity and computational efficiency, four N:M groups are utilized, as increasing the number of groups leads to reduced accuracy. Scaling from four to sixteen groups only raises minimum sparsity by 1.6\%, factoring in SRAM structure. DSST significantly improves accuracy compared to static sparse training, with only a minor decline relative to dense-to-sparse pruning.

\subsection{Acceleration of forward data paths}

\begin{figure}
\centering
\includegraphics[width=0.99\linewidth]{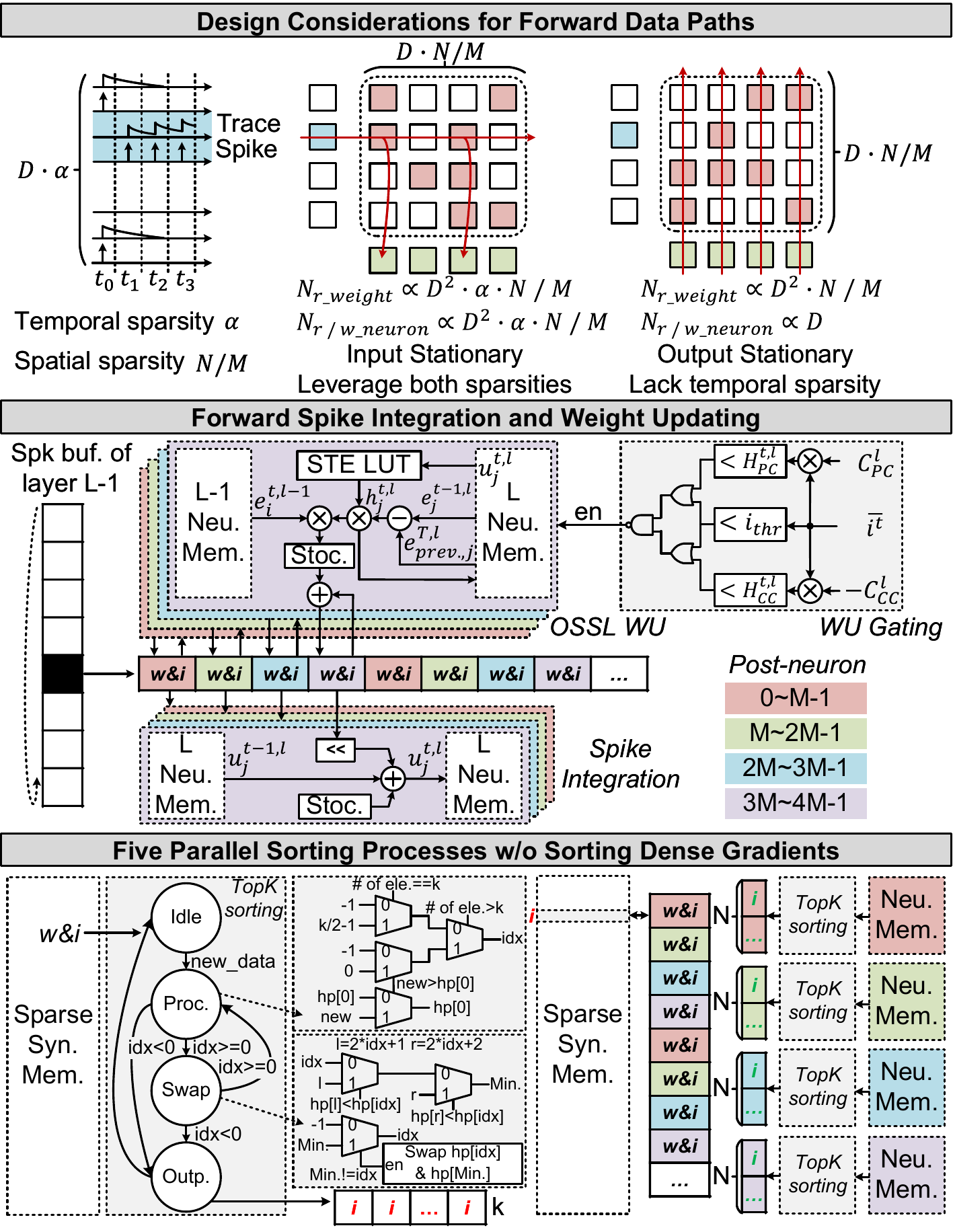}
\caption{
Accelerated forward data paths: Input stationary performs optimally with sparse inputs (top). Parallel layer-wise weight updates and spike integrations (middle). Simultaneous sorting of weights and gradients (bottom).
}
\label{fig:circuits}
\vskip -0.2cm
\end{figure}

Input stationary leverages temporal and spatial sparsity across dual forward paths—traces and spikes (Fig. \ref{fig:circuits}). Four WU and SI PEs operate in parallel over four N:M groups per hidden layer, with all layers pipelined. Post-gradients are written back for DSST sorting, with an straight-through estimator (STE) LUT handling non-derivative spike functions. DSST uses heap sorting with $\mathcal{O}(1)$ space complexity. Each layer has five parallel sorting blocks that identify the k smallest weights from sparse connections and the N largest gradients from M neurons, avoiding full sorting of all M gradients. Pruning and regrowth happen simultaneously by updating indices in the sparse weight memory.

\section{Measurement Results}
\label{sec:measured}

\begin{figure}
\centering
\includegraphics[width=0.99\linewidth]{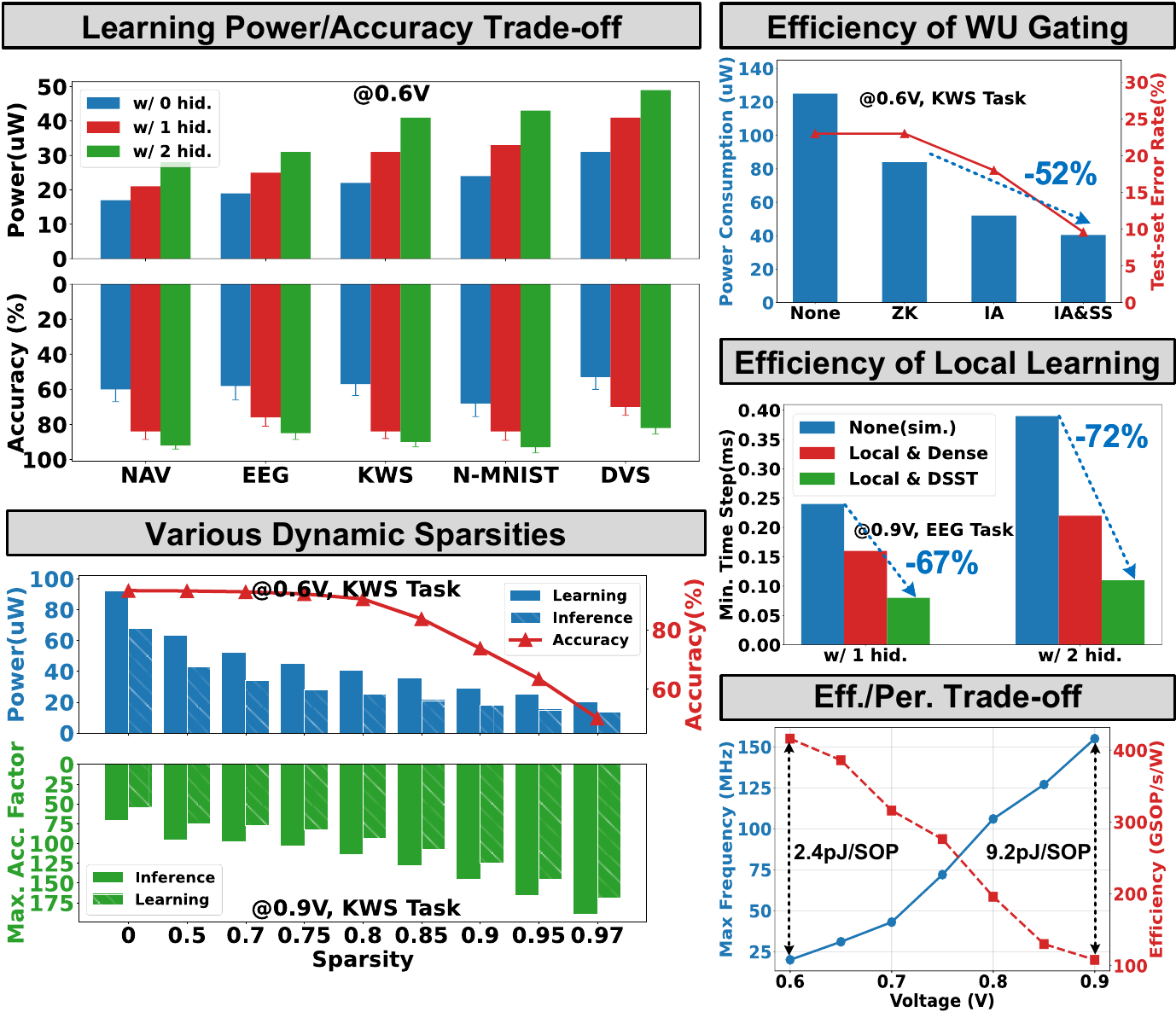}
\caption{Benchmarking and measurement results (average results for 5 chips at 22°C).}
\label{fig:measure}
\end{figure}

End-to-end on-chip learning was evaluated on five temporal tasks, each initialized with random weights and 80\% sparsity (Fig.~\ref{fig:measure}). The hidden-layer bypass mechanism enabled power and accuracy analyses across varying network depths. In combination with DSST, ElfCore’s OSSL effectively learned hierarchical representations while consuming less than 50\,\(\mu\)W for all tasks at 0.6\,V and 20\,MHz.
For the keyword spotting (KWS) task, at 80\% sparsity, DSST reduced learning power by 56\% and inference power by 63\%, while incurring only a 1.8\% drop in accuracy. Moreover, it achieved 1.9\(\times\) faster learning and 1.8\(\times\) faster inference compared to dense training at 0.9\,V and 155\,MHz.
Beyond zero-skipping (ZK), global IA and layer-wise SS gating provided an additional 52\% power reduction, accompanied by an increase in accuracy.
By resolving the WU locking issue, the DSST-based parallel local learning strategy reduced TS length by 67\% for single-hidden-layer networks and by 72\% for two-hidden-layer networks relative to the approach in \cite{vlsi_reckon}, ensuring scalability with increasing network depth.
ElfCore achieves an energy efficiency of 2.4\,pJ/SOP at 0.6\,V.

\section{Conclusion}
\label{sec:conclusion}

\begin{figure}
\centering
\includegraphics[width=0.99\linewidth]{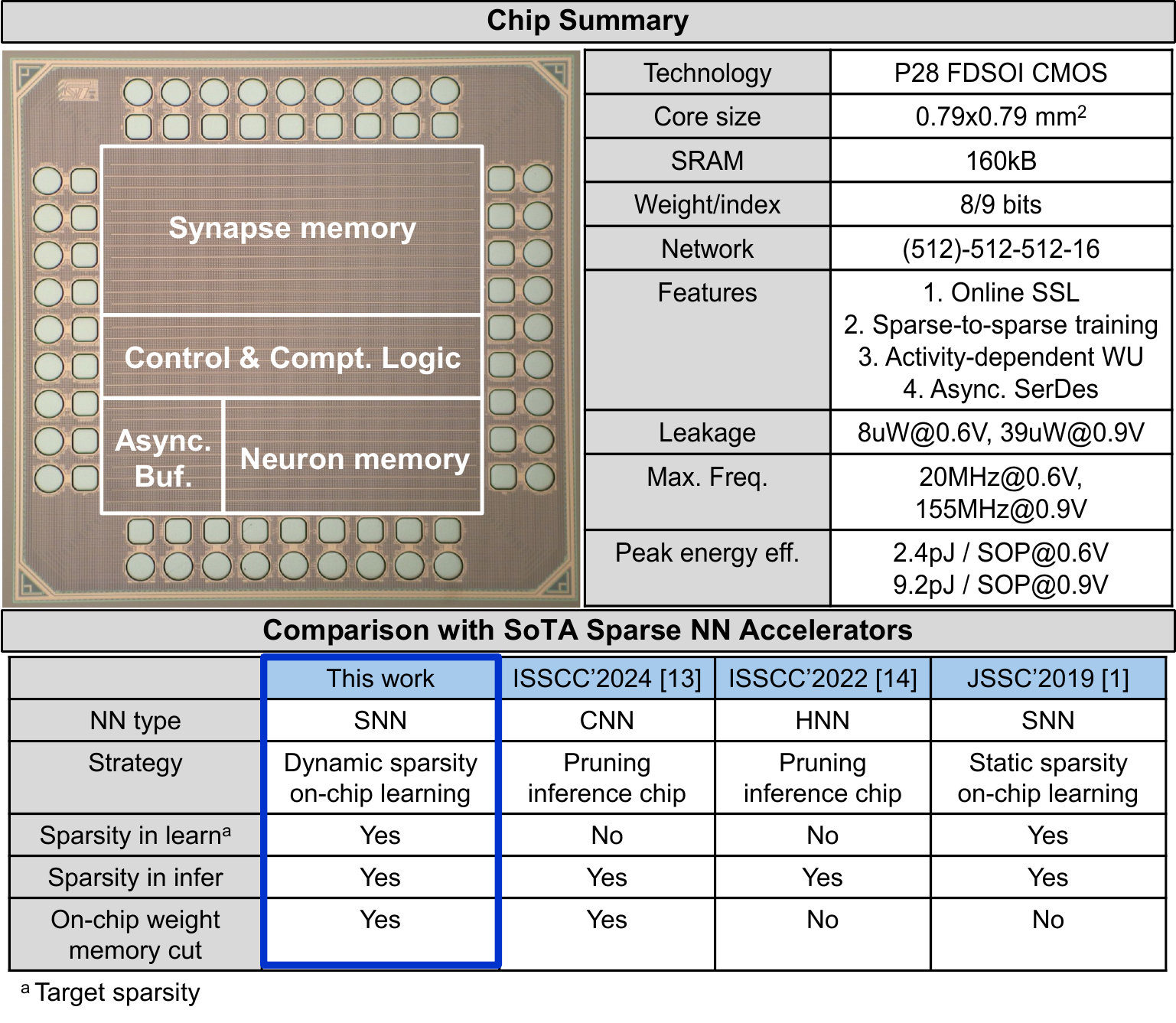}
\caption{Die micrograph (top left), chip summary (top right) and comparison with SoTA sparse neural network accelerators (bottom).}
\label{fig:die}
\vskip -0.2cm
\end{figure}

\begin{table}
\centering
\includegraphics[width=\linewidth]{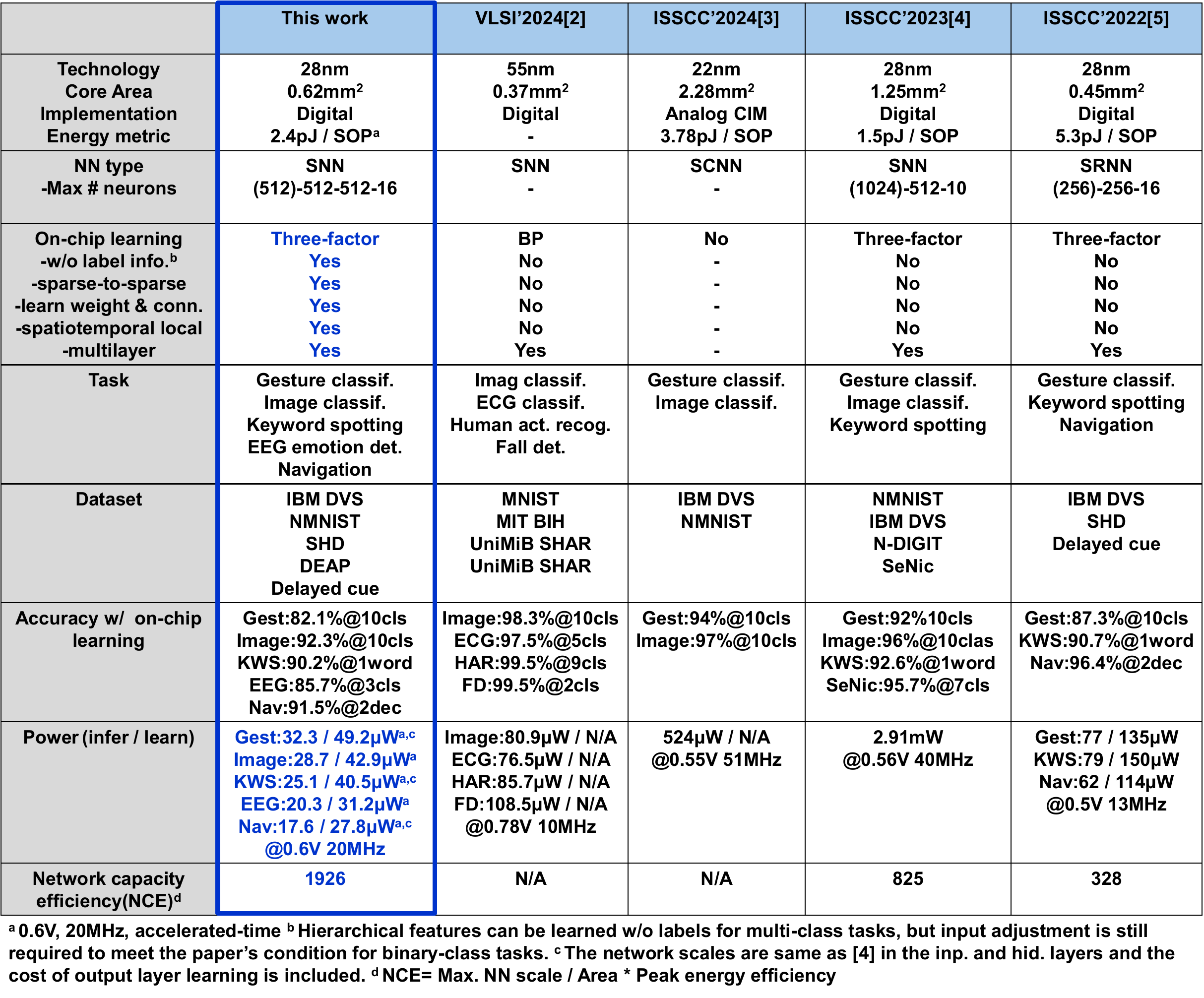}
\caption{Comparison with SoTA SNN processors}
\label{fig:comparison}
\vskip -0.2cm
\end{table}

Unlike earlier attempts that focused on weight sparsity, ElfCore is the first to support end-to-end dynamic sparse-to-sparse training, minimizing the required on-chip weight memory (Fig. \ref{fig:die}). Its unique on-chip three-factor local learning, operating without explicit labels, brings this chip beyond the SoTA SNN processors on same tasks—achieving over 16$\times$ energy savings in inference (vs. \cite{vlsi_tsf}), 3.8$\times$ memory savings at the same network scale, and 4.1$\times$ lower power consumption during learning (vs. \cite{vlsi_reckon}) (Table \ref{fig:comparison}). These features highlight its potential for streaming event data processing. ElfCore is open-source (https://github.com/Zhe-Su/ElfCore.git). 
\nocite{vlsi_pruning, vlsi_hiddenite, vlsi_4096}

% \printbibliography

\section*{Acknowledgment}
This work was supported in part by GA No. 876925 (ANDANTE). We thank the STMicroelectronics R\&D team for their support in the chip design.

\end{document}